\documentclass[twocolumn,english,aps,prl]{revtex4}
\usepackage[T1]{fontenc}
\usepackage[latin9]{inputenc}
\usepackage{amsmath}
\usepackage{graphicx}
\usepackage{amssymb}
\usepackage{esint}

\usepackage{babel}

\begin{document}

\author{Daniel Sank$^1$}
\author{R. Barends$^1$}
\author{Radoslaw C. Bialczak$^1$}
\author{Yu Chen$^1$}
\author{J. Kelly$^1$}
\author{M. Lenander$^1$}
\author{E. Lucero$^1$}
\author{Matteo Mariantoni$^{1,5}$}
\author{M. Neeley$^{1,4}$}
\author{P.J.J. O'Malley$^1$}
\author{A. Vaisencher$^1$}
\author{H. Wang$^{1,2}$}
\author{J. Wenner$^1$}
\author{T.C. White$^1$}
\author{T. Yamamoto$^3$}
\author{Yi Yin$^1$}
\author{A.N. Cleland$^{1,5}$}
\author{John M. Martinis$^{1,5}$}
\email{martinis@physics.ucsb.edu}

\affiliation{$^1$Department of Physics, University of California, Santa Barbara, California 93106-9530, USA}
\affiliation{$^2$Department of Physics, Zhejiang University, Hangzhou 310027, China}
\affiliation{$^3$Green Innovation Research Laboratories, NEC Corporation, Tsukuba, Ibaraki 305-8501, Japan}
\affiliation{$^4$Lincoln Laboratory, Massachusetts Institute of Technology, Lexington, MA 02420-9108}
\affiliation{$^5$California NanoSystems Institute, University of California, Santa Barbara, CA 93106-9530, USA}

\date{\today}

\title{Surface spin fluctuations probed with flux noise and coherence \\ in Josephson phase qubits}

\begin{abstract}
We measure the dependence of qubit phase coherence and surface spin induced flux noise on inductor loop geometry. While wider inductor traces change neither the flux noise power spectrum nor the qubit dephasing time, increased inductance leads to a simultaneous increase in both. Using our protocol for measuring low frequency flux noise, we make a direct comparison between the flux noise spectrum and qubit phase decay, finding agreement within 10\% of theory. The dependence of the measured flux noise on inductor geometry is consistent with a noise source correlation length between 6 and 400\,$\mu$m.
\end{abstract}
\maketitle


Superconducting qubits \cite{ClarkeWilhelm:SuperconductingQuantumBits} are rapidly approaching the requirements needed for fault tolerant quantum computation, with recent experiments demonstrating a set of information storage, logic gates, and coherence improvements \cite{DiCarlo:ThreeQubitEntanglement,Neeley:ThreeQubitEntanglement,Wang:NOONStates,Paik:3DTransmon}. However, in recent demonstrations of both multi-qubit gates with phase qubits \cite{Bialczak:Tomography,Mariantoni:VonNeumann} and error correction with transmons \cite{Reed:ThreeQubitErrorCorrection}, the fidelity of the quantum process was limited by individual qubit dephasing times $T_2$. This dephasing results from low frequency fluctuations in the magnetic flux used to tune the devices' $|0\rangle \rightarrow |1\rangle$ transition frequency $f_{10}$. Traditionally, flux, transmon, and quantronium qubits have addressed this problem by idling at ``sweet spots'' where the first order sensitivity of device frequency to bias flux, $df_{10}/d\Phi$, is zero \cite{Chiorescu:FluxQubit,Koch:Transmon,Vion:Quantronium}. This approach has limitations, however, as qubits have to be moved from their idle points to participate in multi-qubit quantum gates. Furthermore, this approach to the flux noise problem constrains the qubit design space, as devices must be carefully engineered to have a sweet spot.

Anomalous low frequency flux noise in superconducting devices has a long history in both experiment and theory \cite{Wellstood:FluxNoise,Choi:MIGs,Faoro:FluxNoise,Koch:FluxNoise,Sendelbach:SQUIDMillikelvinTemperatures,Sendelbach:SQUIDSusceptometer}. Experiments with SQUIDs have found that the flux noise power spectral density $S_{\Phi}(f)$, which scales with frequency approximately as $1/f$, is insensitive to device size and materials; in fact no strong dependence on any device parameter has been found. Recent experiments have identified spins on the metal surfaces as the noise source \cite{Sendelbach:SQUIDMillikelvinTemperatures,Sendelbach:SQUIDSusceptometer}, but its exact microscopic mechanism remains unknown. Experiments with qubits have measured the dependence of dephasing on the bias point, characterized by $df_{10}/d\Phi$ \cite{Yoshihara:FluxNoise,Ithier:Decoherence,Kauyanagi:FluxQubitDecoherence}, and have made direct \cite{Bialczak:1fNoise} and indirect \cite{Bylander:FluxNoiseSpectroscopy} measurements of the noise spectrum, but no experiment has carefully compared dephasing times with a direct measurement of the low frequency flux noise, nor has any experiment demonstrated a means of improving $T_2$ independently of the bias point.

In this Letter, we present measurements of dephasing times and flux noise in phase qubits designed to improve $T_2$. In a circuit with increased inductance, we find an improvement in phase coherence at all bias points. We introduce a protocol for directly measuring low frequency flux noise and use this protocol to make a clear and quantitative comparison between qubit dephasing and the measured noise spectrum. With this protocol, we also report the first strong dependence of the flux noise on device geometry.

In order to motivate our redesigned qubits, we first review dephasing theory. Dephasing is caused by random fluctuations in the qubit's transition frequency $f_{10}$, characterized by a spectral density $S_{f_{10}}(f)$. For a Ramsey fringe experiment, the theoretical prediction for the time dependent decay of the qubit population probability is \cite{Martinis:BiasNoise} \begin{equation}
p(t) = \exp \left[ -\frac{(2\pi)^2}{2} t^2 \int_{f_{\textrm{m}}} ^{\infty} S_{f_{10}}(f)\, \left( \frac{\sin(\pi f t)}{(\pi f t)}\right)^2 df \right],
\label{eq:ramseyIntegral}
\end{equation}
where $f_{\textrm{m}}$ is a lower cutoff frequency equal to the inverse of the total experiment time \cite{footnote:gaussianNoise}. In  phase qubits, frequency fluctuations are dominated by flux noise with a nearly $1/f$ spectral density, \mbox{$S_{\Phi}(f) = S^*_{\Phi}/f^{\alpha}$}, \mbox{$\alpha \approx 1$} \cite{Bialczak:1fNoise}, resulting in a frequency noise spectral density \mbox{$S_{f_{10}}(f) = (df_{10}/d\Phi)^2 S^*_{\Phi}/f^{\alpha}$}. Inserting this into Eq.\,(\ref{eq:ramseyIntegral}), performing the integral for the case $\alpha=1$, and adding the contribution from energy loss ($T_1$), yields \begin{equation}
\ln[ p(t) ] =-\frac{t}{2T_{1}}-\frac{(2\pi)^2}{2} t^{2} \left(\frac{df_{10}}{d\Phi} \right)^2  S_{\Phi}^* \ln\left(\frac{0.4}{f_{\textrm{m}}t}\right).
\label{eq:ramseyCurve}
\end{equation}
The log factor is weakly dependent on $t$ and has a numerical value of $\sim$24 in the experimental range \mbox{$10 \, \textrm{ns} < t < 400 \, \textrm{ns}$} \cite{_supplementary}. Defining the dephasing time $T_2^*$ by the equation \begin{equation}
\ln [p(t)] = -t/2T_1 - (t/T_2^*)^2 ,
\label{eq:T2Definition}
\end{equation}
we find \begin{equation}
T_2^* \propto S_{\Phi}^{*-1/2} \left( \frac{df_{10}}{d\Phi} \right)^{-1} = S_{\Phi}^{*-1/2} L \left( \frac{df_{10}}{dI} \right)^{-1} , \label{eq:T2formula}
\end{equation}
where $L$ is the inductance of the qubit loop and $I$ is the loop current. Evidently $T_2$ can be increased by decreasing the sensitivity $df_{10}/dI$, but this incurs a decrease in the qubit's nonlinearity $\Delta$, defined as \mbox{$\Delta \equiv f_{21}-f_{10}$}. Nonlinearity is a critical figure of merit, as greater nonlinearity allows for shorter control pulses and therefore more quantum gates during the qubit's lifetime \cite{Lucero:highFidelityGates}. We therefore re-express $df_{10}/dI$ in terms of $\Delta$ according to $df_{10}/dI \propto \Delta ^{3/4}$ \cite{Martinis:BiasNoise}, and substitute into Eq.\,(\ref{eq:T2formula}), yielding \begin{equation}
\ln(T_2^*)=-\frac{3}{4}\ln(\Delta)+\ln (L)-\frac{1}{2}\ln(S_{\Phi}^{*})+K.
\label{Eq:T2vsNonlin}\end{equation}
Here $K$ depends on other device parameters, such as capacitance and junction critical current, which control the device's operating frequency range. As we do not wish to change the device frequency, $K$ must remain fixed. Similarly, as we wish to improve $T_2$ without sacrificing nonlinearity, we regard $\Delta$ as a scaling parameter. There remain two ways to improve $T_2^*$: raise $L$ or decrease $S_{\Phi}^{*}$.

\begin{figure}
\begin{centering}
\includegraphics[width=9cm]{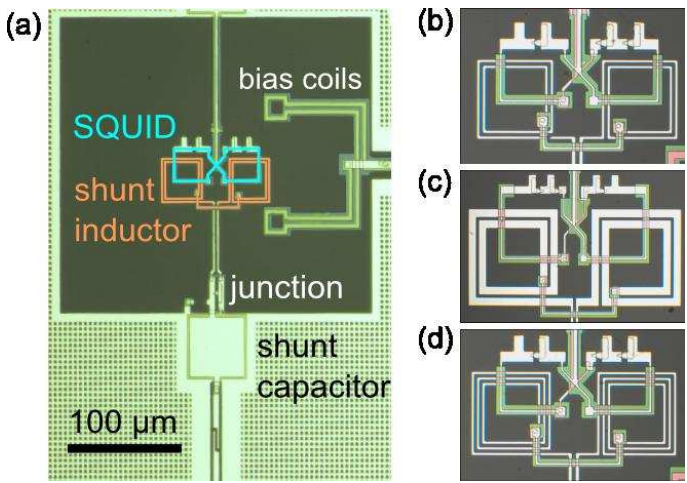} 
\par\end{centering}
\caption{(color online) Phase qubit devices used in the experiments. Panel (a) shows the readout SQUID, inductor coils, bias coils, and shunt capacitor; panels (b), (c), and (d) show close-ups of the inductors and SQUIDs. (b) Standard qubit coils with trace widths $w = 1.5$ $\mu$m and $L$ = 710 pH. (c) Wide trace coils with $w = 6$ $\mu$m and $L$ = 720 pH. (d) High inductance coils with $w = 1.5$ $\mu$m and $L$ = 1330 pH.}
\label{Figure:devices} 
\end{figure}

Each of these two methods was implemented though a redesign of the qubit inductor loop. The devices are pictured in Fig.\,\ref{Figure:devices}. The standard phase qubit \cite{Martinis:SuperconductingPhaseQubits,Neeley:TransformedDissipation} is shown in Fig.\,\ref{Figure:devices}a, with a close-up of the inductor coils in Fig.\,\ref{Figure:devices}b. Coils of the first redesign are shown in Fig.\,\ref{Figure:devices}c. Theories attributing flux noise to localized, uncorrelated magnetic moments of unpaired electron spins on the metal surfaces predict that $S_{\Phi}^*$ scales as $R/W$ where $R$ is the radius of the metal loop and $W$ is its width \cite{Bialczak:1fNoise,Sendelbach:SQUIDMillikelvinTemperatures}. Therefore, to lower $S_{\Phi}^*$, we increased the trace widths from the standard 1.5 $\mu$m to 6.0 $\mu$m. The second redesign is shown in Fig.\,\ref{Figure:devices}d. In this device, we increase the inductance of the loop by adding turns, increasing from 710 pH to 1330 pH. Increased inductance should reduce noise currents driven by noise flux and lead to better $T_2^*$ as per Eq.\,(\ref{Eq:T2vsNonlin}).

\begin{figure}
\begin{centering}
\includegraphics[width=9cm]{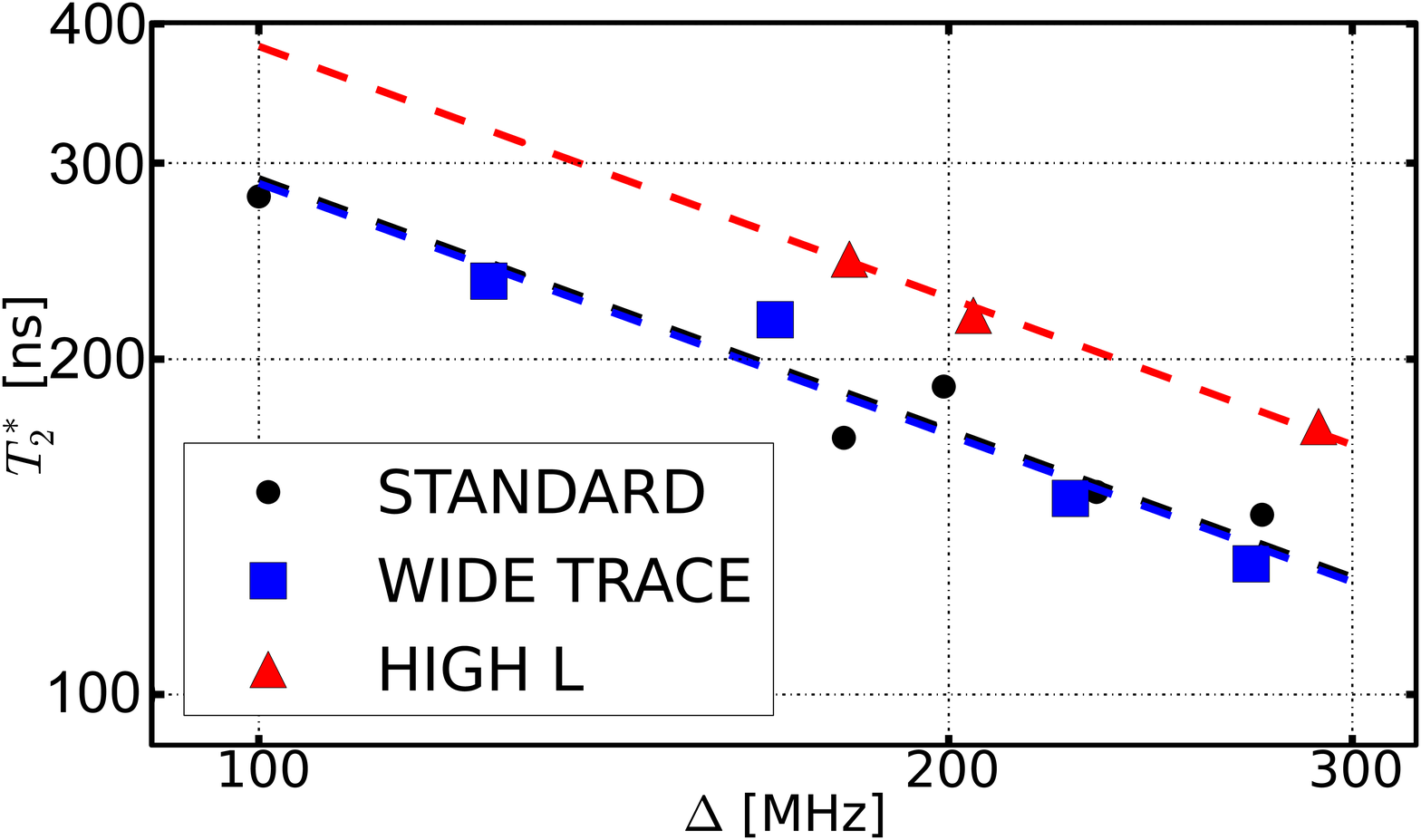} 
\par\end{centering}
\caption{(color online) Plot of Ramsey decay time $T_2^*$ vs. nonlinearity $\Delta$ for each type of device. Each point comes from a fit to Eq.\,(\ref{eq:T2Definition}). For each bias flux, $T_1$ and $\Delta$ are measured in separate experiments. Dashed fit lines are drawn with the predicted slope of $-3/4$ in order to compare magnitudes. The fit lines for the standard and wide trace designs fall on top of one another. The high inductance devices show an increase in $T_2^*$ by a factor of 1.3. Confidence bounds, typically 2\% of data values, cannot be seen on this scale. }
\label{Figure:T2vsNonlinearity} 
\end{figure}

In the first set of experiments we measured $T_2^*$ at several flux bias points that varied $\Delta$. The exponential energy decay time $T_1$ and nonlinearity $\Delta$ are first measured in separate experiments \cite{Martinis:RabiOscillations,Lucero:DRAG}. We then measure a Ramsey decay curve and fit the data to Eq.\,(\ref{eq:T2Definition}) so that the contribution from $T_1$ is appropriately removed in obtaining the Gaussian decay constant $T_2^*$.

The results shown in Fig.\,\ref{Figure:T2vsNonlinearity} reveal that the dependence of the extracted values of $T_2^*$ on $\Delta$ is consistent with the expected -3/4 power law. This indicates that $T_2^*$ scales inversely with the qubit frequency sensitivity, as predicted by Eq.\,(\ref{eq:T2formula}). The wide trace devices show no difference in $T_2^*$ from the standard design over the measured range of operating points. Because the inductance of the wide trace design is equal to the inductance of the standard design, the lack of change in $T_2^*$ suggests that $S_{\Phi}^*$ was not lowered by the increase in trace width, consistent with previous results in SQUIDs \cite{Wellstood:FluxNoise}. The high inductance device showed an increase in $T_2^*$ by 30\% relative to the standard design at all bias points. This is our first main result: increased inductance improves the qubit's dephasing time independently of the bias point, and without sacrificing any other figure of merit.

The disparity between the 1.30$\times$ measured improvement and the 1.87$\times$ improvement predicted by Eq.\,(\ref{Eq:T2vsNonlin}), based on the designed increase in $L$, suggests that the high inductance device had an increase in flux noise. According to Eq.\,(\ref{Eq:T2vsNonlin}) the relative improvement in $T_2^*$ is expected to be \begin{equation}
\frac{T_2^{*'}}{T_2^*} = \frac{L'\sqrt{S_{\Phi}^*}} {L\sqrt{S_{\Phi}^{*'}}} .
\label{eq:relativeT2}
\end{equation}
For the measured $T_2^*$ increase of 1.30 and the designed inductance increase of 1.87, Eq. (\ref{eq:relativeT2})
predicts an increase in noise amplitude of $\sqrt{S_{\Phi}^{*'}/S_{\Phi}^*} = 1.4$.

\begin{figure}
\begin{centering}
\includegraphics[width=8cm]{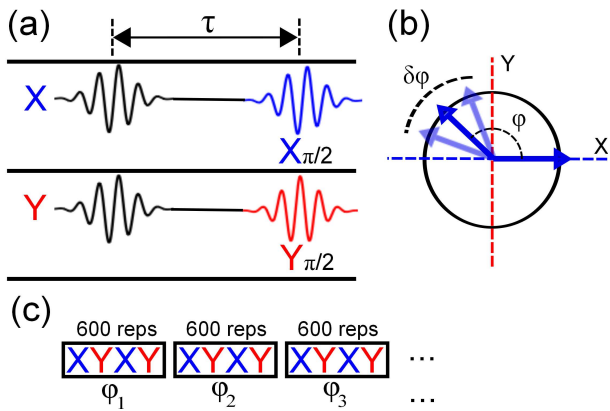} 
\par\end{centering}
\caption{(color online) The Ramsey tomography oscilloscope (RTO) protocol. (a) Four Ramsey sequences are performed, two with the final $\pi/2$ pulse about the $\pm X$ axes, and two with the final $\pi/2$ pulse about the $\pm Y$ axes. The precession time for both sequences is fixed at $\tau=~100\textrm{ns}$. (b) The combined $X$ axis and $Y$ axis sequences constitute tomography of the state in the x-y plane after precession by $\tau$, thereby measuring the angle $\varphi$ traversed by the state. (c) Several hundred subsequent $X$ and $Y$ measurements are grouped together to get an average value for $\varphi$. The entire process is repeated once per second for several hours to build up a time series $\delta f(t)=\delta\varphi(t)/2\pi\tau$.}
\label{Figure:pulseSequence} 
\end{figure}

We next test these predictions for the relative noise changes against a direct measurement of the flux noise. The Ramsey fringe measurement is sensitive to the time dependent fluctuations of the qubit frequency averaged over the entire data acquisition time, and therefore measures \emph{integrals} of the noise spectral density. To \emph{directly} probe $S_{\Phi}(f)$, one can instead track the qubit frequency in real time and then produce a spectrum with Fourier transform methods. A previous experiment with phase qubits implemented this idea by measuring the time dependent fluctuations of the position of the qubit resonance peaks \cite{Bialczak:1fNoise}. The present experiment refines this idea: instead of measuring $f_{10}$ spectroscopically, we use the free precession of the qubit state in the equator of the Bloch sphere. The angle traversed by the state in a fixed time $\tau$ is $\varphi=2\pi f_{10}\tau$. By choosing $\tau$ and measuring $\varphi$ through partial tomography we obtain $f_{10}$. Repetition of this measurement for several hours produces a time series $f_{10}(t)$. This measurement protocol, which we call the Ramsey tomography oscilloscope (RTO), is illustrated in Fig.\,\ref{Figure:pulseSequence}. The RTO requires little calibration and can be implemented in any qubit system.

\begin{figure}
\begin{centering}
\includegraphics[width=9cm]{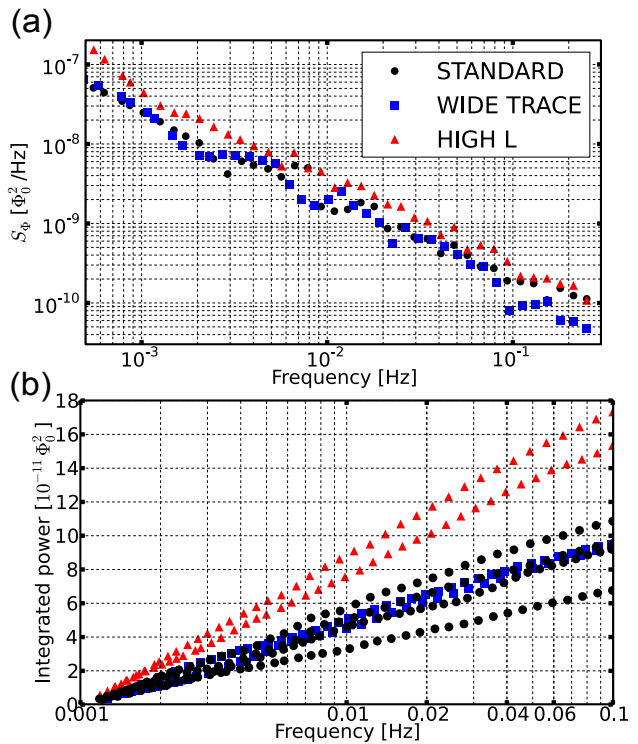} 
\par\end{centering}
\caption{(color online) Power spectra for each inductor design. (a) Typical spectra for each type of device. (b) Spectral power integrated over the range 0.001 Hz to 0.1 Hz. In (a) we plot only one spectrum for each device design for clarity in the plot. In (b) we show several curves, measured at several bias points, in devices all fabricated on the same chip.}
\label{Figure:spectra}
\end{figure}

The time series $f_{10}(t)$ is converted to a periodogram using the discrete Fourier transform, and the result is appropriately normalized to produce the power spectral density of frequency noise $S_{f_{10}}(f)$ \cite{_supplementary}. The sensitivity $df_{10}/d\Phi$ is measured separately, and the flux noise spectral density is obtained from the frequency noise spectral density according to $S_\Phi(f) = (df_{10}/d\Phi)^{-2}S_{f_{10}}(f)$. Results are shown in Fig. \ref{Figure:spectra}. Note in Fig.\,\ref{Figure:spectra}a that the spectra contain little statistical noise, in contrast to the data of Ref.\,\cite{Bialczak:1fNoise} where the measured power spectra wandered by almost an order of magnitude within each decade of frequency.

The power spectra for each type of device scale as $f^{-1.1}$ over the measured band and have extrapolated amplitudes at 1 Hz between 3.5 and 5.0 $\mu\Phi_0/\sqrt{\textrm{Hz}}$. The spectra in Fig.\,\ref{Figure:spectra}a directly show that the high inductance device has more flux noise than the other two devices. Still, a clearer picture is obtained if we integrate the power spectral density to produce a plot of cumulative power. To avoid complications from aliasing and spectral leakage from DC, we omit the upper and lower edges of the measured frequency band, integrating only over the range $10^{-3}$ to $10^{-1}$ Hz where the measurement is the least susceptible to these effects \cite{Harris:Windows}. From the integrated power shown in Fig.\,\ref{Figure:spectra}b, it is clear that the wide trace and normal designs have the same level of noise, while the high inductance device shows a 1.7$\times$ increase in noise power.

For the wide trace devices, the lack of change in noise level relative to the standard devices is consistent with the results of the Ramsey fringe experiments in which these two designs had the same values of $T_2^*$. For the high inductance device, the $\sqrt{1.7}=1.3$-fold increase in noise amplitude found in the RTO is in approximate agreement with the results from the Ramsey fringe experiments and Eq.\,(\ref{eq:relativeT2}), which predict an increase of 1.4.

We have also used the RTO to measure cross correlation in the noises of two phase qubits separated by 500 $\mu$m on the same chip. We find that the measured correlation is no greater than that found for two independently simulated $1/f$ noise signals \cite{_supplementary}.

The scaling of the noise power with inductance, and the lack of scaling with trace aspect ratio, have implications for models of the flux noise. The lack of change with aspect ratio is incompatible with models predicting $S_{\Phi}^* \propto R/W$, such as the independent surface spins proposed in Ref.\,\cite{Bialczak:1fNoise}, or single electrons interacting via the superconducting condensate as proposed in Ref.\,\cite{Faoro:FluxNoise}. This may indicate a correlation length larger than the trace width. If the noise sources were correlated over the entire length of the coil, we would expect the noise power to scale quadratically with the number of coil turns, while for uncorrelated sources it would scale linearly. The high inductance devices had 50\% greater number of turns than the standard devices, and we found a noise power increase of 70\%, closer to the uncorrelated case. The near linear scaling is also evidence that the noise sources produce local fields; non-local fields would allow each source to couple spin into multiple turns of the coil, making the scaling of noise power with number of turns super-linear. A possible interpretation of all of these results is that the noise sources are localized, but have a correlation length that is greater than the 6 $\mu$m trace width and smaller than the 400 $\mu$m total length of the high inductance coil.

Thus far we have discussed the \emph{relative} dephasing times and flux noise levels. It remains to check whether the the \emph{absolute} measured noise levels accounts for the observed dephasing times. This is done by comparing the value of $S_{\Phi}^{*\textrm{Ramsey}}$ extracted from a fit to the Ramsey data with the value $S_{\Phi}^{*\textrm{RTO}}$ found directly using the RTO. Note that previous experiments with qubits have been unable to make this comparison because an accurate measurement of the flux noise was unavailable. As described by Eq.\,(\ref{eq:ramseyIntegral}) and the surrounding discussion, the exact form of the Ramsey fit function depends on the scaling power $\alpha$. In particular, it would be inappropriate to fit the Ramsey data to Eq.\,(\ref{eq:ramseyCurve}); that equation was derived under the assumption that $\alpha$ is equal to 1, whereas we found in the RTO that $\alpha$ is closer to 1.1. To take the scaling power into account properly, we numerically evaluate the integral in Eq.\,(\ref{eq:ramseyIntegral}) for $\alpha = 1.1$, and use the resulting fit function to extract $S_{\Phi}^{*\textrm{Ramsey}}$ \cite{_supplementary}. For all devices and bias points we find that the values of $S_{\Phi}^{*\textrm{Ramsey}}$ extracted in this way agree with $S_{\Phi}^{*\textrm{RTO}}$ to within $10\%$. We emphasize that properly accounting for the \emph{measured} slope of the noise spectral density when fitting the Ramsey curves is essential in obtaining agreement between the Ramsey and RTO experiments. For example, if we take $\alpha=1$ we find \mbox{$S_{\Phi}^{*\textrm{Ramsey}} \approx 4 \, S_{\Phi}^{*\textrm{RTO}}$}.

In conclusion, we have found that increased loop inductance is a viable means to improve superconducting qubit phase coherence independently of bias point. We have introduced the RTO protocol for directly measuring flux noise, and used this protocol to make a direct comparison between measured flux noise and qubit dephasing times. There we find that the noise level extracted from Ramsey decay agrees with the direct measurement only when the exact slope of the noise spectrum is considered. From another point of view, this sensitivity means that Ramsey decay times can only be accurately predicted when the scaling power of the noise is known. Using the RTO we compared the flux noise in devices with widened traces and increased number of inductor turns. With no change found in the wide traces, but a nearly linear increase in noise with number of turns, we conclude that the noise sources may be correlated over distances greater than the 6 $\mu$m trace width, but smaller than the 400 $\mu$m length of the inductor coils. Clearly, the correlation length of the flux noise sources is a key parameter that should be studied in further experiments.

We thank Robert McDermott for discussions on flux noise in superconductors. This work was supported by Intelligence Advanced Research Projects Activity (IARPA) under ARO award W911NF-08-1-0336 and under Army Research Office (ARO) award W911NF-09-1-0375. M. M. acknowledges support from an Elings Postdoctoral Fellowship. Devices were made at the University of California Santa Barbara Nanofabrication Facility, a part of the NSF-funded National Nanotechnology Infrastructure Network.

\bibliographystyle{apsrev}

\begin{thebibliography}{31}
\bibitem[{foo()}]{footnote:gaussianNoise}
\bibinfo{note}{This formula was derived in \cite{Martinis:BiasNoise} as an
  approximation for the case of small noise. It is exact for Gaussian noise.}
\end{thebibliography}

\end{document}


\author{Daniel Sank$^1$}
\author{R. Barends$^1$}
\author{Radoslaw C. Bialczak$^1$}
\author{Yu Chen$^1$}
\author{J. Kelly$^1$}
\author{M. Lenander$^1$}
\author{E. Lucero$^1$}
\author{Matteo Mariantoni$^{1,5}$}
\author{M. Neeley$^{1,4}$}
\author{P.J.J. O'Malley$^1$}
\author{A. Vaisencher$^1$}
\author{H. Wang$^{1,2}$}
\author{J. Wenner$^1$}
\author{T.C. White$^1$}
\author{T. Yamamoto$^3$}
\author{Yi Yin$^1$}
\author{A.N. Cleland$^{1,5}$}
\author{John M. Martinis$^{1,5}$}
\email{martinis@physics.ucsb.edu}

\affiliation{$^1$Department of Physics, University of California, Santa Barbara, California 93106-9530, USA}
\affiliation{$^2$Department of Physics, Zhejiang University, Hangzhou 310027, China}
\affiliation{$^3$Green Innovation Research Laboratories, NEC Corporation, Tsukuba, Ibaraki 305-8501, Japan}
\affiliation{$^4$Lincoln Laboratory, Massachusetts Institute of Technology, Lexington, MA 02420-9108}
\affiliation{$^5$California NanoSystems Institute, University of California, Santa Barbara, CA 93106-9530, USA}

\date{\today}

\title{Supplementary material for Surface spin fluctuations probed with \\ flux noise and coherence in Josephson phase qubits}

\begin{abstract}
Here we present experimental details on the Ramsey Tomography Oscilloscope (RTO) protocol and details of the calculations used to extract the flux noise magnitude from Ramsey decay data.
\end{abstract}
\maketitle

\section{Ramsey Tomography Oscilloscope}

Here we describe the data processing procedure for the Ramsey Tomography Oscilloscope (RTO). We found that careful signal processing was important in reducing statistical noise in the power spectra generated by the RTO. The bandwidth of the RTO measurement is set fundamentally by the rate at which the qubit can be measured and reset. In our case this would allow ideally 10,000 quantum measurements per second. With our current asynchronous control software this limit could not be reached while simultaneously tracking the time at which each measurement occurred. Maximum data rate with accurate time stamping was achieved with 2,400 quantum measurements per second, 600 of each of the four tomography sequences. Averaging the 600 measurements together produced one frequency measurement per second. This set the bandwidth of the experiment to be 0.5 Hz due to the Nyquist criterion.

\begin{figure}
\begin{centering}
\includegraphics[width=9cm]{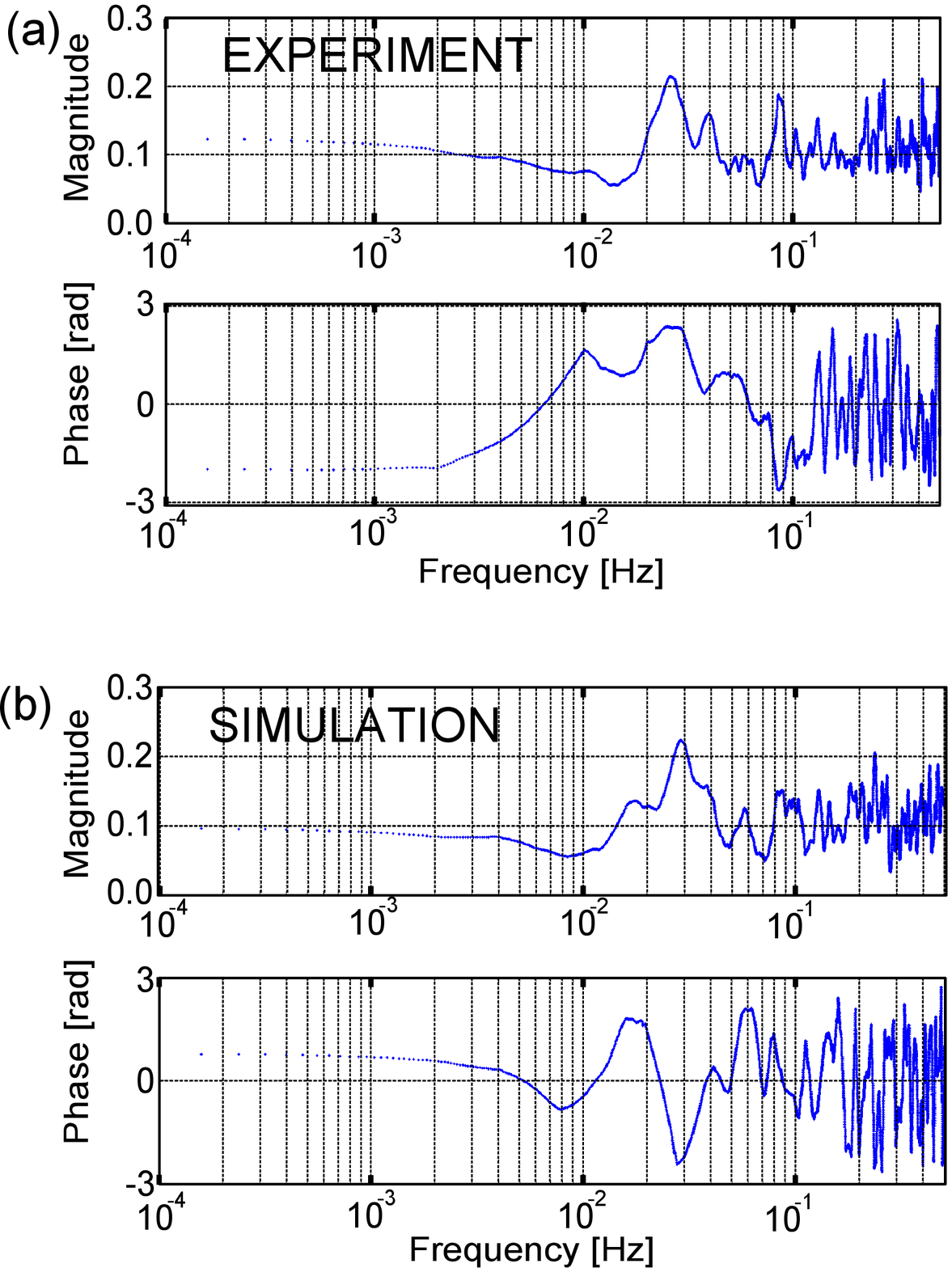} 
\par\end{centering}
\caption{Cross spectra. (a) Cross spectrum measured using the RTO. (b) Cross spectrum computed from two independently simulated $1/f$ noise signals.}
\label{Figure:crossSpectrum}
\end{figure}

Data was typically acquired for eight to ten hours, yielding between 28,000 and 36,000 points in the time series. Power spectra are computed as follows. First, the time series is divided into four or five non-overlapping sections. We compute the power spectrum of each section separately and average them together at the end of the procedure. To eliminate uncorrelated quantum measurement shot noise, we use an interleaving procedure on each section. Each section is split into two interleaved time series, $f_1(n)$ and $f_2(n)$ ($n$ is the discrete time index). These series are multiplied by Hann windows and the discrete Fourier transforms $F_1(k)$ and $F_2(k)$ are computed ($k$ is a frequency bin index). We form the product $F_1(k)F_2^*(k)$, average neighboring bins together using a Gaussian weight function with full width at half maximum (FWHM) of 20 bins, and take the magnitude to obtain the periodogram $P(k)$. Next, the periodogram is multiplied by a factor 1/0.375 to correct for the loss of incoherent (noise) power caused by application of the Hann window \cite{Harris:Windows}. The periodogram is then smoothed by averaging neighboring frequency bins with a Gaussian weight function with a variable FWHM scaling quadratically from 1 bin at the low end of the frequency band to 20 bins at the high end. The power spectrum $S(f)$ is then computed from the periodogram according to \begin{equation}
S(f)=\frac{2T}{(N/2)^2}P(k=fT) \end{equation}
where $T$ is the total length of time represented by the section of the time series, and $N$ is the number of points in the section. Finally, spectra generated from each section are averaged together.

\section{Cross Correlation}

In order to check that the flux noise we measured was generated locally to each device, we used the RTO to measure cross correlation of the noise signals generated in two devices separated by 500 $\mu$m on the same chip. Time series of the two devices' resonance frequencies were measured using the RTO, and the cross correlation was computed. Results are shown in Fig.\, \ref{Figure:crossSpectrum}. Although there are frequencies at which the cross correlation amplitude is as high as 0.3, this must be compared against the cross correlation computed for two independently simulated noise signals. We find that the cross correlation of two independently simulated $1/f$ noise signals show very similar peak structure to the data, indicating that the noises within the two qubits are no more correlated than independent noise. This result agrees with the finding in Ref.\,\cite{Bialczak:Tomography}, where it was inferred from quantum state tomography performed on two coupled qubits that dephasing in each qubit was uncorrelated. We note that the absence of a low frequency roll-off in the RTO data indicates that the low frequency flux noise is correlated on time scales exceeding the length of data acquisition. For this reason it is unsurprising that residual cross-correlation was found in both the data and the simulation.

\section{Comparison of RTO and Ramsey Decay}

\begin{figure}
\begin{centering}
\includegraphics[width=9cm]{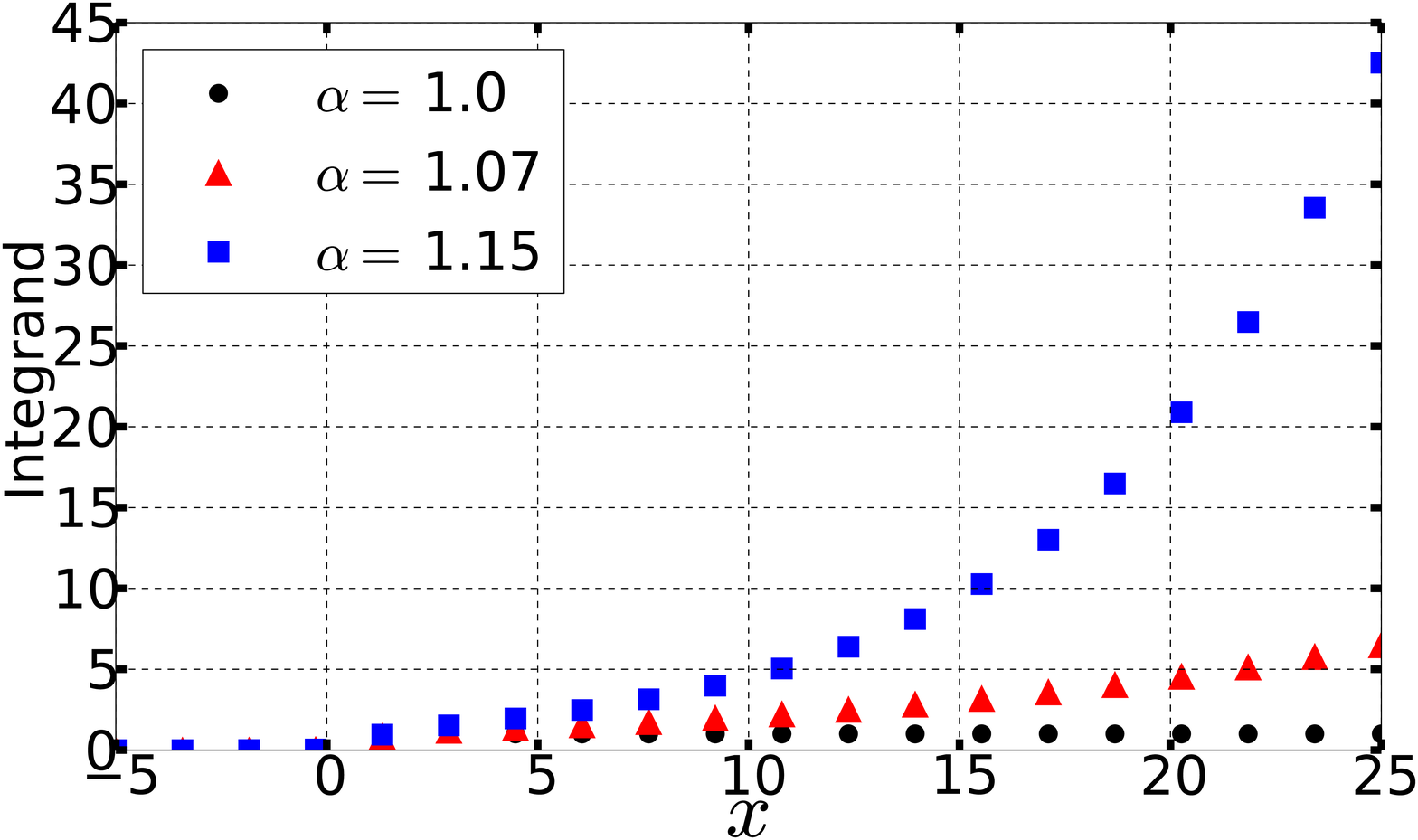} 
\par\end{centering}
\caption{(color online) The integrand of $I$. The curves are well behaved over the entire integration range.}
\label{Figure:ramseyIntegrands}
\end{figure}

\begin{figure}
\begin{centering}
\includegraphics[width=9cm]{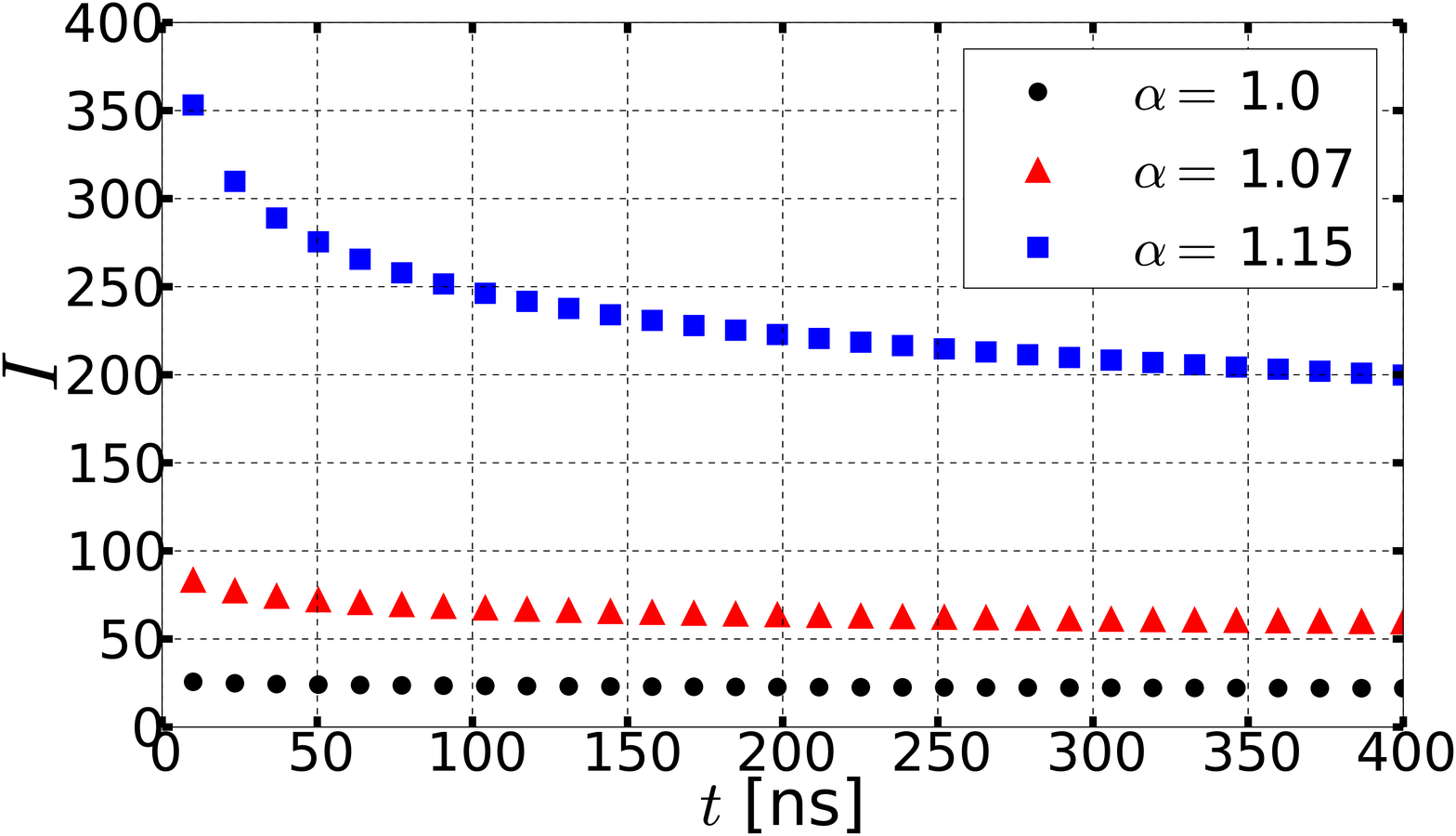} 
\par\end{centering}
\caption{(color online) The integral $I$ evaluated versus $t$ for several values of $\alpha$. Note the strong sensitivity to $\alpha$; as $\alpha$ goes from 1.0 to 1.15, a $15\%$ change, the integral increases by a factor of $\sim$10.}
\label{Figure:ramseyIntegrals}
\end{figure}

We wish to fit our Ramsey decay data to the theoretical curve given by Eq.\,(1) in the main text, which, for the case where the flux noise is $S_{\Phi}(f)=S_{\Phi}^*/f^{\alpha}$, is \begin{equation}
p(t) = \exp \left[ -\frac{(2\pi)^2}{2} \left( \frac{df_{10}}{d\Phi}\right)^2 S_{\Phi}^* \, t^{1+\alpha} \int_{f_{\textrm{m}}}^{\infty}  \frac{\textrm{sin}(\pi z)^2}{(\pi z)^2} \frac{dz}{z^{\alpha}} \right]
\label{eq:ramseyFormula}
\end{equation}
Here $\textrm{sinc}(x)\equiv \sin(x)/x$, $f_{\textrm{m}} \approx 1$ hour  and $t$ is in the range 0 to 400 ns \footnote{The lower cutoff frequency $f_{\textrm{m}}=1/\textrm{hour}$ arises from the manner in which we acquire the data. Because we use a projective quantum measurement to read the state of the qubit, we must repeat the experiment for each value of $t$ many times to reduce statistical noise. Rather than average each point in $t$ sequentially, we spread the averaging of each point over the entire trace acquisition period. This results in better averaging of the noise signal as explained in Ref.\,\cite{VanHarlingen:CriticalCurrentDecoherence}.}. In order to do this we need to evaluate the integral \begin{equation}
I = \int_{f_{\textrm{m}}t}^{\infty} \frac{\sin(\pi z)^2}{(\pi z)^2}\frac{dz}{z^{\alpha}} .
\end{equation}
We compute the integral numerically. Since $f_{\textrm{m}}t$ is on the order of $10^{-12}$, the lower limit of integration is a very small positive number and the integrand is diverging at the lower limit. On the other hand, the integrand oscillates for $z>1$. The integral is therefore unfit for numerical analysis in its current form as it has both divergent and oscillatory behavior. The problem is mitigated by the change of variables $x\equiv -\ln(z)$ which yields \begin{equation}
I = \int_{-\infty}^{-\ln(f_{\textrm{m}}t)} \frac{\sin(\pi e^{-x})^2} {(\pi e^{-x})^2}\frac{dx}{e^{x(\alpha-1)}} \end{equation}
The integrand is now well conditioned over the whole integration range. Plots of this integrand for several values of $\alpha$ are shown in Fig.\,\ref{Figure:ramseyIntegrands}. Note that an upper cutoff in the frequency integral would translate to a lower cutoff in the integral over $x$. Because of the logarithmic scale combined with the very small value of the integrand for values of $x \leq 5$, ignoring a possible upper cutoff greater than 1\,MHz incurs negligible error.

\begin{figure}
\begin{centering}
\includegraphics[width=9cm]{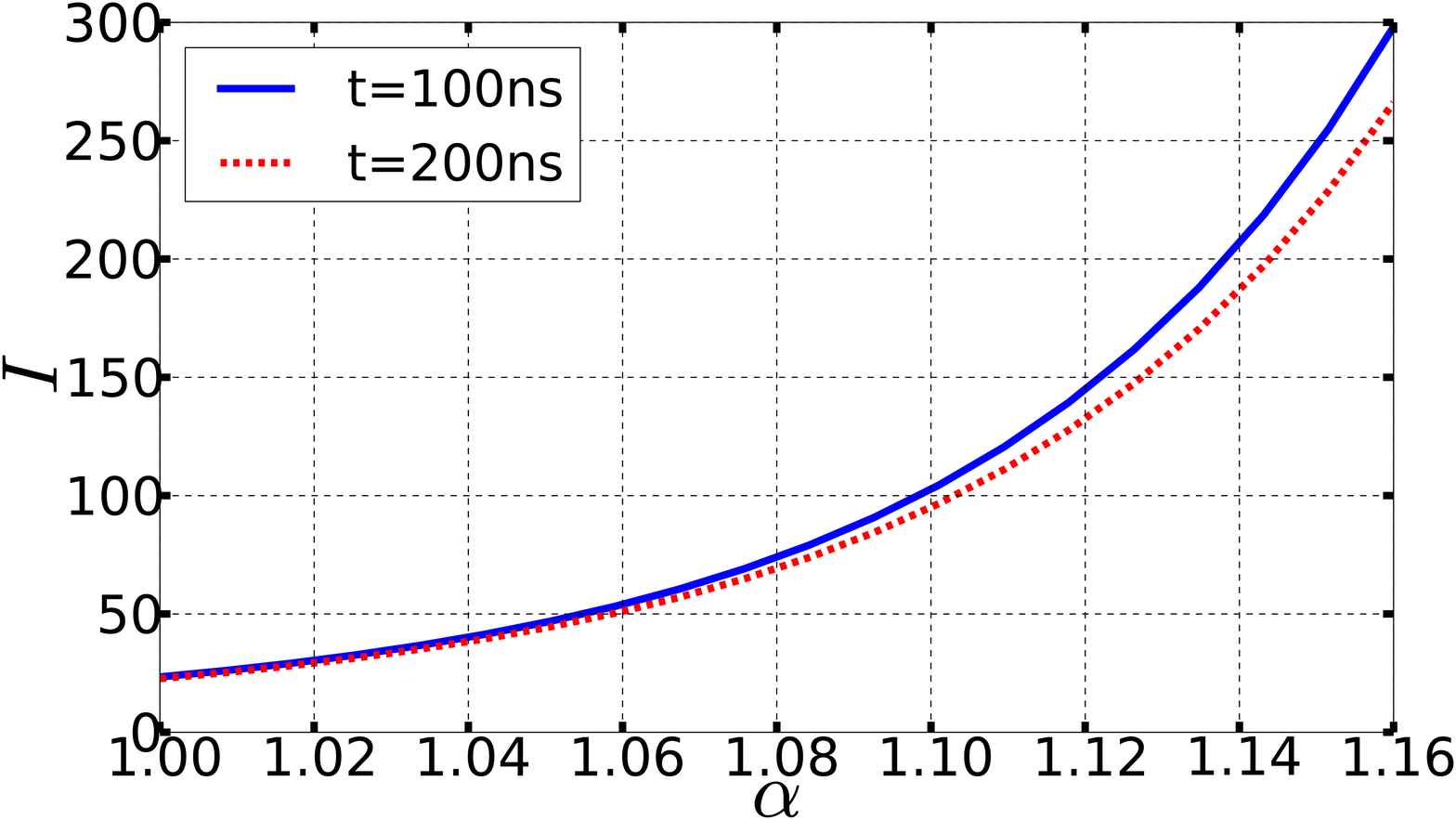} 
\par\end{centering}
\caption{(color online) The integral $I$ evaluated as a function of $\alpha$ for several values of $t$. Note the strong dependence on $\alpha$.}
\label{Figure:ramseyIntegralsVsAlpha}
\end{figure}

We perform the integral $I$ for 50 values of $t$ in the experimental range 0 to 400 ns, and for several values of $\alpha$ near 1. Results of the integration as a function of $t$ are shown in Fig.\,\ref{Figure:ramseyIntegrals}. We also show $I$ as a function of $\alpha$ for two fixed values of $t$ in Fig.\,\ref{Figure:ramseyIntegralsVsAlpha}. From these curves we construct interpolating functions and use them to fit our measured Ramsey decay data to Eq.\,(\ref{eq:ramseyFormula}). Note particularly in Fig.\,\ref{Figure:ramseyIntegralsVsAlpha} the strong dependence of the noise integral on $\alpha$. It is because of this strong dependence that we are able to accurately determine which value of $\alpha$ gives the best agreement with the power spectra measured directly using the RTO, as described in the main text.

\bibliographystyle{apsrev}